\documentclass[10pt,a4paper,twocolumn]{article}
\usepackage{adjustbox}
\usepackage{float}
\usepackage{tikz}
\usepackage{amsmath}
\usepackage{amssymb}
\usepackage{pgfplots}
\usepackage[numbers]{natbib}
\usepackage{xcolor}
\usepackage{changepage}
\usepackage{multirow} 
\usepackage{lscape}
\usepackage[font=small,labelfont=bf]{caption}
\usepackage[justification=centering]{caption}
\usepackage{enumerate}
\usepackage{graphicx,dblfloatfix}
\usepackage{comment}
\graphicspath{ {image/} }
\usepackage{verbatim}
\usepackage{longtable}
\usepackage{lipsum}
\usepackage{comment}
\usepackage{subcaption}
\usepackage{setspace}
\usepackage{abstract}
\usepackage[margin=1.0in]{geometry}

\usepackage{lineno}  
\usepackage{authblk}

\pgfplotsset{compat=1.17}

\begin{document}
	\title{Hungry Professors? Decision Biases Are Less Widespread than Previously Thought\thanks{ \emph{\scriptsize{We are very grateful to Daisy Riccio Buqi for the outstanding help with the data preparation. We also thank the Ethics Committee of the Faculty of Business and Economics (HEC) of the University of Lausanne for their helpful comments.}}}}
	
	% Use letters for affiliations, numbers to show equal authorship (if applicable) and to indicate the corresponding author
	%\author{Katja Bergonzoli, Laurent Bieri, Dominic Rohner, Christian Zehnder\\University of Lausanne}
	\author[a]{Katja Bergonzoli}
	\author[a]{Laurent Bieri} 
	\author[a,b]{Dominic Rohner}
	\author[a]{Christian Zehnder}
	
	\affil[a]{\footnotesize University of Lausanne}
	\affil[b]{\footnotesize CEPR}

	\date{\today}
	%\doi{\url{www.pnas.org/cgi/doi/10.1073/pnas.XXXXXXXXXX}}
	
	%\pgfplotsset{compat=1.17}
	
	\twocolumn[
	\begin{@twocolumnfalse}
		\maketitle
		%\singlespacing
		\begin{abstract}{\footnotesize In many situations people make sequences of similar, but unrelated decisions. Such decision sequences are prevalent in many important contexts including judicial judgments, loan approvals, college admissions, and athletic competitions. A growing literature claims that decisions in such sequences may be severely biased because decision outcomes seem to be systematically affected by the scheduling. In particular, it has been argued that mental depletion leads to harsher decisions before food breaks and that the ``law of small numbers'' induces decisions to be negatively auto-correlated (i.e. favorable decisions are followed by unfavorable ones and vice versa). These findings have attracted much academic and media attention and it has been suspected that they may only represent the ``tip of the iceberg''. However, voices of caution point out that existing studies may suffer from serious limitations, because the decision order is not randomly determined, other influencing factors are hard to exclude, or direct evidence for the underlying mechanisms is not available. We exploit a large-scale natural experiment in a context in which the previous literature would predict the presence of scheduling biases. Specifically, we investigate whether the grades of randomly scheduled oral exams in Law School depend on the position of the exam in the sequence. Our rich data enables us to filter-out student, professor, day, and course-specific features. Our results contradict the previous findings and suggest that caution is advised when generalizing from previous studies for policy advice.
				\vspace{0.05in}
				
				\noindent\textbf{\footnotesize Keywords: Biased decisions, Grading in academia, Law studies}
			}
		\end{abstract}\vspace{0.5cm}\thispagestyle{empty}
	\end{@twocolumnfalse}
	]
	\saythanks
	
	%\ifthenelse{\boolean{shortarticle}}{\ifthenelse{\boolean{singlecolumn}}{\abscontentformatted}{\abscontent}}{}
	
	% If your first paragraph (i.e. with the \dropcap) contains a list environment (quote, quotation, theorem, definition, enumerate, itemize...), the line after the list may have some extra indentation. If this is the case, add \parshape=0 to the end of the list environment.
	%\dropcap{I}ntro.
	%\subsubsection*{Research Question \& Literature Review}
	
	Life is full of situations, in which people make long sequences of similar, but unrelated decisions. Typical examples include HR professionals who conduct a series of job interviews, credit analysts who evaluate a list of loan applications, teachers who grade exams, or call center agents who respond to customer complaints. Over the last decade, such choice environments have received growing attention in the literature, because several studies have reported systematic and severe biases in decision sequences. In particular, researchers have claimed that decision outcomes not only depend on the relevant underlying information, but also on the position in the decision schedule.
	
	A seminal article \cite{danziger2011extraneous} studies a sample of parole decisions by judges in Israel and finds that the share of favorable rulings falls gradually from 65 percent at the beginning of a decision session to nearly zero before a food break and then peaks again when the judge is well-fed. As these spectacular results raise serious questions about legal fairness, this study had a tremendous impact on the discipline and eventually opened up a whole new field of research. The mechanisms envisaged involve (chemical) processes linked to hunger and low glucose levels \cite{petersen2014social,orquin2016meta,vicario2018effect} and psychological processes related to ``decision fatigue'' or ``mental depletion'' \cite{muraven2000self,vohs2018making}. The broad applicability of these mechanisms suggests that these findings might represent just the tip of the iceberg and that similar biases may occur throughout a variety a high-stake decisions in politics, business and science.
	
	Given the potentially wide-ranging policy implications of this study, it is key to make sure that its conclusions are robust and carry over to different contexts with a similar structure. A potentially important concern about \cite{danziger2011extraneous} is that the cases over which judges preside may not be scheduled randomly. If ``easier'' cases with higher success chances are for some reason scheduled earlier, this sorting could drive the results. An independent study \cite{weinshall2011overlooked} claims that the scheduling is indeed non-random. While the reply of the authors of the original article \cite{danziger2011reply} rules out one particular type of selection bias (represented vs non-represented prisoners), various other reasons for non-random scheduling can be imagined (e.g., higher parole likelihood in some prisons than in others, more capable and better-organized attorneys arriving earlier in the morning, etc). Simulations \cite{glockner2016irrational} show that the findings could also be accounted for by a pattern where unfavorable decisions take longer than favorable ones. Moreover, given that cognitive fatigue and boredom vary over the day and week \cite{mark2014bored,sievertsen2016cognitive}, it is hard to disentangle the judges' biases from the attorneys' quality of pleading which may also vary substantially depending on the time of the day.
	
	Beyond this pioneering work, there exist other, more recent studies on the impact of scheduling on decisions. For example, negative auto-correlation in decisions have been reported for refugee asylum court decisions, loan application reviews, and Major League Baseball umpire pitch calls \cite{chen2016decision}. It is argued that these findings are most likely explained by the law of small numbers and the gambler’s fallacy, i.e. the tendency that decision makers underestimate the probability of randomly occurring sequential streaks. In addition, our work also relates to the broader literature studying potential biases resulting from decision fatigue. Such effects have been shown in studies on voting choices \cite{augenblick2016ballot}, analyst forecasts \cite{hirshleifer2019decision}, medical decisions \cite{linder2014time,allan2019clinical}, and moral judgments in the lab \cite{timmons2019moral}.
	
	If the aforementioned evidence were to generalize to other contexts, the implications would be dramatic. However, the robustness of these results remains unclear for two reasons. First, as mentioned earlier, some studies suffer from methodological problems, because scheduling is not random, the time per decision is not kept constant, or it is impossible to identify which of several interacting parties creates the observed bias. Second, most studies do not allow to fully pin down the mechanisms underlying the results. This second point is crucial, because it is the mechanism that determines whether and to what extent a finding can be extrapolated to other contexts. To emphasize the potentially broad and general importance of its findings, the existing literature tends to point to very general channels such as hunger and mental depletion \cite{danziger2011extraneous} or the law of small numbers \cite{chen2016decision}. In this study, we exploit a large-scale natural experiment in a context where one would expect to see the previously described decision biases if the underlying mechanisms were correctly identified. In addition, we have access to very rich data that allow us to rule out the above mentioned confounds. 
	
	In particular, we study the impact of scheduling (time, order of appearance, before or after a food break) for the grading of oral exams of Bachelor and Master students at the Law School of the University of Lausanne in Switzerland, drawing on over 14,000 observations. Importantly, the schedule (running order) in our setting is completely exogenous. In particular, the exams are organized centrally by the administration and the assignment of individual exam slots can plausibly be described as random. %are assigned in a , with the order being typically randomized / alphabetical.
	Further, since each student passes several exams and each professor evaluates several students, we are able to include in our multiple regression analysis both student and professor ``fixed effects'' (i.e. student and professor specific constant terms) that filter out unobserved characteristics, such as a given student being stellar in all disciplines or a given professor grading on average harshly throughout. Given that several exams take place in parallel, we are also able to include another battery of temporal fixed effects (i.e. a different constant term for each exam date), which control for e.g. excessive heat on a given day or for being the first day of the week. %Finally, data on student performances in written exams also allow us to disentangle whether any potential biases are driven by professor biases or student performance. For example, if there was a large grade decline over the day in oral exams but not in written ones, we would be able to attribute this to professor bias rather than student performance. 
	These features of the data allow for an arguably much cleaner and more robust statistical investigation of the impact of scheduling on decision biases than in other contexts.
	
	\section*{Materials and Methods}
	
	The statistical analysis of the current paper uses new original data from the Law School of the University of Lausanne, covering all oral exams that took place from 2018 to 2021, for the Winter, Spring, and Summer sessions and for both Bachelor and Master programs. We exclude exams graded by more than one Professor, and group exams (where at least two students took the exam together).
	
	\begin{figure}[ht!]
		\centering
		\begin{subfigure}[a]{0.4\textwidth}
			\centering
			\includegraphics[width=\textwidth]{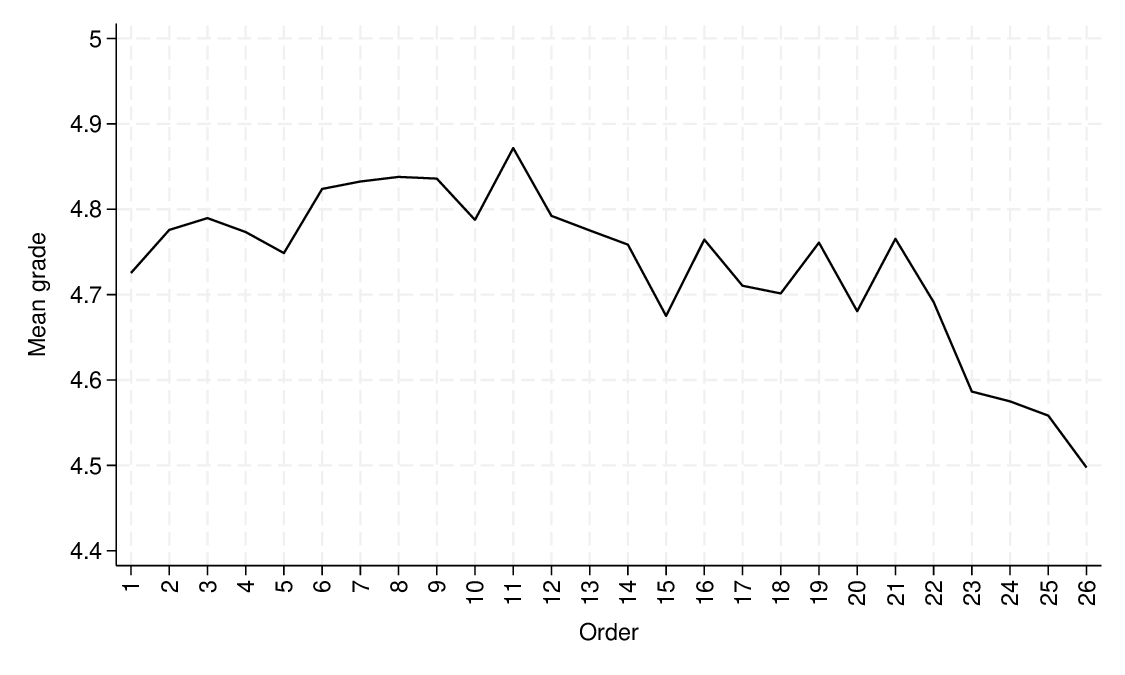}
			\caption{Average Grade by Order of Passage}
			\label{gig:figure1a}
		\end{subfigure}
		\newline
		\vspace{0.5cm}
		\begin{subfigure}[b]{0.4\textwidth}
			\centering
			\includegraphics[width=\textwidth]{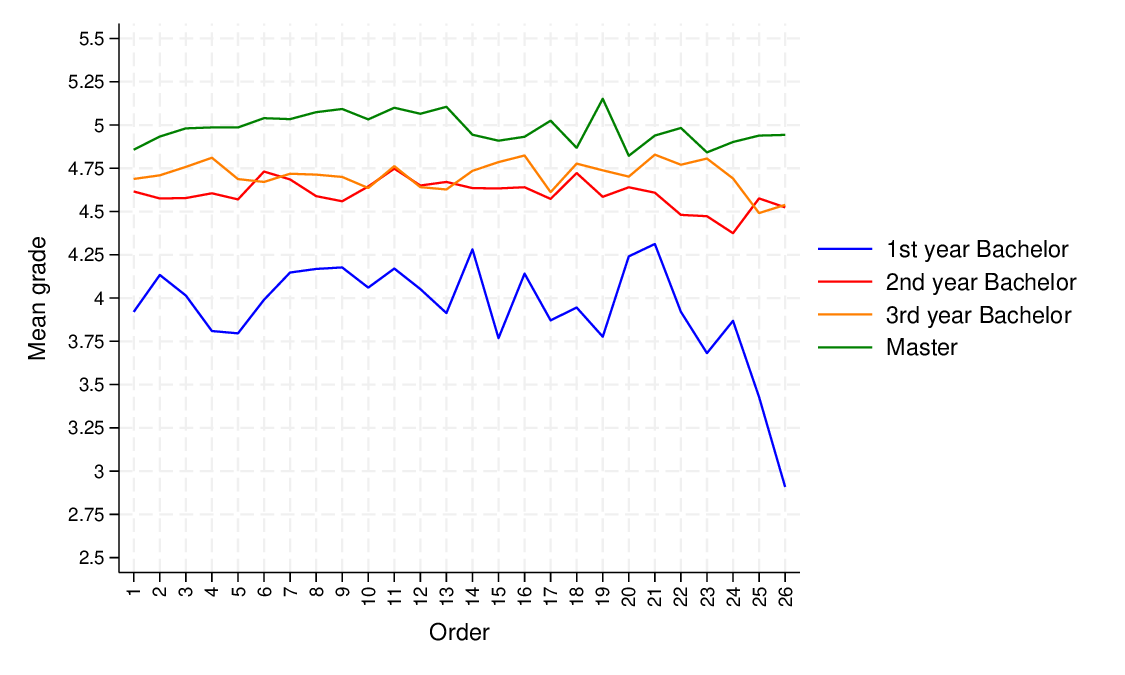}
			\caption{Average Grade by Order of Passage by Level of Studies}\label{fig:figure1b}
		\end{subfigure}
		\newline
		\vspace{0.5cm}
		\begin{subfigure}[c]{0.4\textwidth}
			\centering
			\includegraphics[width=\textwidth]{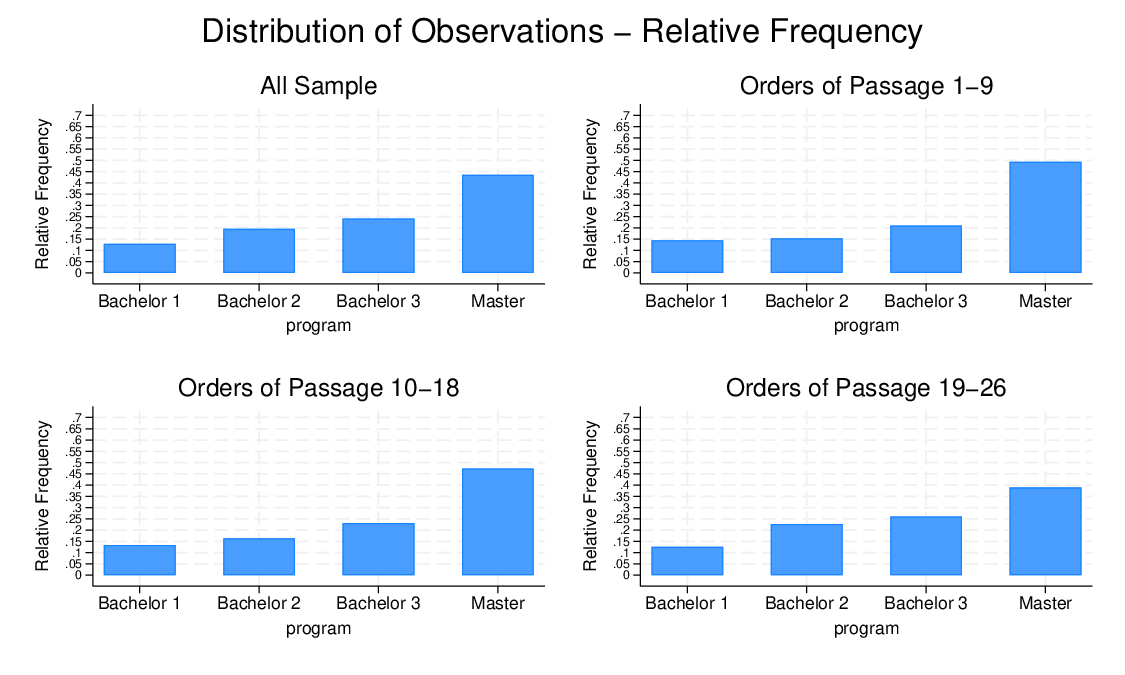}
			\caption{Distribution of Observations}\label{fig:figure1c}
		\end{subfigure}
		\begin{subfigure}[a]{0.4\textwidth}
			\centering
			\includegraphics[width=\textwidth]{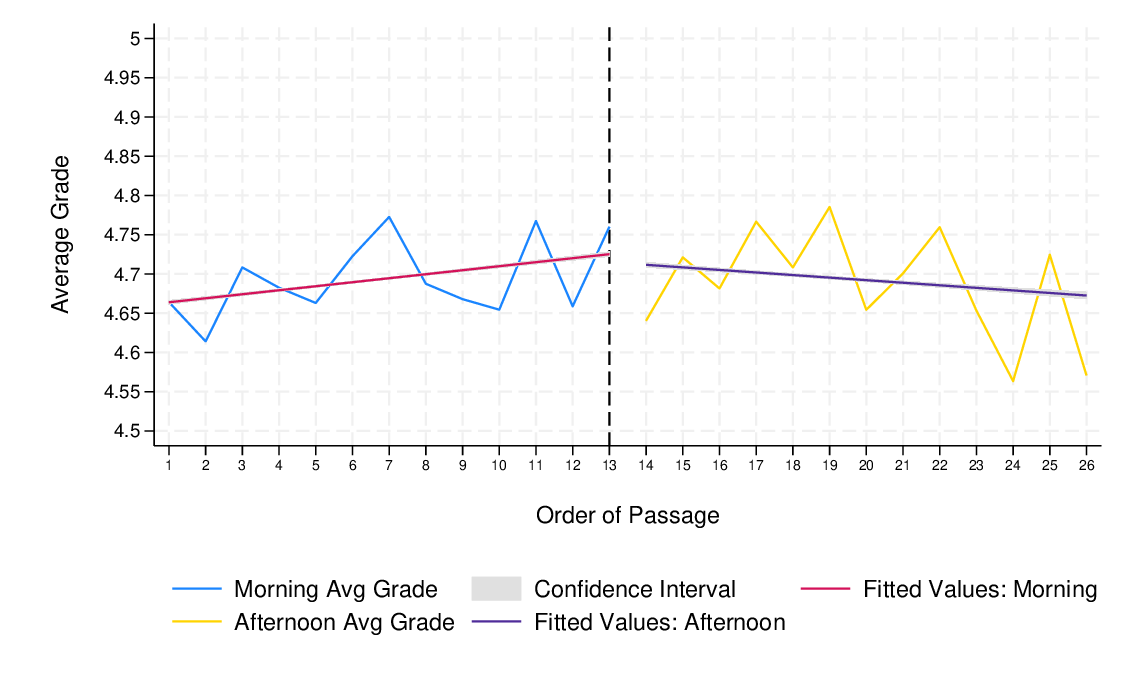}
			\caption{Mean grades before and after the break (Graphically)}
			\label{fig:figure2a}
		\end{subfigure}
		\newline
		\vspace{0.5cm}
		\caption{Descriptive Figures}\label{fig:figure1}
	\end{figure}

	Appendix Table \ref{tab:AT1} presents the descriptive summary statistics. The main dependent variable of interest is $Grade$. This variable can take the values $\{1,1.25,1.5,1.75,2,...,6\}$ and represents the grade given to a student by a professor for an oral exam at a specific date and time. Note that in Switzerland, the lowest grade is $1$ and the best grade is $6$ with $4$ being the passing grade. The grade of 0 also exists, but it is reserved for unjustified absences. We therefore removed observations with a grade of 0 from the sample (because this grade does not reflect the professor's evaluation of the student).
	
	The key explanatory variable \textit{Order} is the running order of students for a given date, exam, and professor. The running order is calculated for each specific exam, professor and date, not taking into account unjustified absences and continuing the counting throughout the day (i.e. breaks are not taken into account). In our main specifications, we use the running order as a linear variable. However, as robustness checks, we also present alternative specifications that include indicator variables for different time slots during the day: \textit{7-8am, 9-10am, 11am-12pm, 1-2pm, 3-4pm, 5-6pm}. Each variable captures and intervals of two hours (for example, the variable \textit{7-8am} takes the value 1 if an exam takes place between 7am and strictly before 9am).
	
	Another variable of interest is the average grade of the previous three students, allowing us to study for the presence of auto-correlation (our specific procedure was inpired by \cite{gilovich1985hot}).

	%Depending on the specification, we use different sets of control variables. The variables \textit{Date}, \textit{Time} are respectively the date and the time of the exam. The variables \textit{Start time}, \textit{End time}, \textit{Time length} and \textit{Break length} are calculated for a given date, exam, and professor. %Note that \textit{Break length} can count for both lunch and no lunch breaks. 
	
	In the Appendix we perform a robustness analysis, where we draw on further information. We observe the gender (female or male) of professors and of students, allowing us to code bilateral variables capturing constellations where both professor and student are female, or both are male. As our sample covers the period of the COVID-19 pandemic, some of the exams took place online on Zoom, which is also captured by an indicator variable. We also draw on information on \textit{No lunch break}, \textit{First exam of the day}, and \textit{Last exam of the day}, which are binary variables. 
	
	The variable \textit{Program} captures whether a given exam is a first-year Bachelor exam, second-year Bachelor exam, third-year Bachelor exam, or Master exam. The variable \textit{Exam} has a separate value for each different course graded with an oral exam (say, ``Introduction to Law'', ``Business Law'' etc.). Last but not least, the \textit{Student id} and \textit{Professors id} variables are IDs created specifically for this study and do not correspond to any actual ID system.
	
	%\textit{Morning} takes the value 1 for exams happening in the morning; for exams without a lunch break, \textit{Morning} is equal to one for exams that took place before 2 p.m.; for exams with a lunch break, \textit{Morning} is equal to one for exams that took place before the lunch break. 
	
	Our main specification is a fixed effects regression with clustered standard errors at the professor level (there are 76 professors in our sample):
	
	%\begin{figure}[tbhp]%[bt!]
	%\begin{align*}
	%
	%
	%\end{align*}
	%\end{figure}
	
	\begin{align*}\label{eqn:reg1}
		Grade_{i,s,p,t} = & \alpha + \beta * Order_{i,s,p,t} +  \gamma * C_{i,s,p,t} \\
		& + \delta_i + \zeta_s + \eta_p + \theta_t + \epsilon_{i,s,p,t},
	\end{align*}
	
	\noindent where $Grade_{i,s,p,t}$: for exam $i$ and student $s$, with professor $p$, on date $t$. $Order_{i,s,p,t}$: Running order, $C_{i,s,p,t}$: Controls (as listed above), $\delta_i,\zeta_s,\eta_p,\theta_t$: Respectively exam, student, professor, date fixed effects.
	
	%\begin{comment}
	%\begin{figure*}[bt!]%[tbhp]%[bt!]
	%\begin{align*}
	%Grade_{i,s,p,t} = & \alpha + \beta * X_{i,s,p,t} +  \gamma * C_{i,s,p,t} \\
	%& + Male\_Student_s*Male\_Prof_p + Male\_Student_s*Female\_Prof_p \\
	%& + Female\_Student_s*Male\_Prof_p \numberthis \label{eqn:reg2} \\
	%& + X_{i,s,p,t}*Female\_Student_s + X_{i,s,p,t}*Female\_Prof_p \\
	%& + X_{i,s,p,t}*Female\_Prof_p*Female\_Student_s \\
	%& + \delta_i + \zeta_s + \eta_p + \theta_t + \epsilon_{i,s,p,t}.
	%\end{align*}
	%\end{figure*}
	
	%\begin{itemize}
	%	\item $G_{i,s,p,t}$: Grade for student s, for exam i, with professor i, in session t
	
	%	\item $X_{i,s,p,t,m}$: Order of passing, Hour of passing (dummies variables)
	
	%	\item $ C_{i,s,p,t}$: Controls (zoom, morning, exam length, day of the week, lunch break or not, break length, first exam of the day, last exam of the day, first exam after lunch break, date)
	
	%	\item $\delta_i,\zeta_s,\eta_p,\theta_t$: Respectively exam, student, professor, session-year FEs
	
	%\end{itemize}
	
	%\textbf{Panel }
	%\begin{itemize}
	%	\item The pannel I used: \\
	%	\textit{xtset prof\_id datetime}
	%	\medskip
	%\end{itemize}
	%\textbf{Cluster}
	%\begin{itemize}
	%	\item Cluster at the \textit{prof\_id} level
	%\end{itemize}
	
	%\textbf{Regression}
	%\begin{itemize}
	%	\item \textit{xtreg grade x controls, fe cluster(prof\_id)}
	%\end{itemize}
	
	%\end{comment}

\begin{table}[H]
	\centering
	\caption{Main results on passing order and times\label{tab:T1}}
	\adjustbox{max width=0.5\textwidth}{
		\begin{tabular}{l*{6}{c}}
			\hline\hline
			
			&\multicolumn{1}{c}{(1)}&\multicolumn{1}{c}{(2)}&\multicolumn{1}{c}{(3)}&\multicolumn{1}{c}{(4)}&\multicolumn{1}{c}{(5)}&\multicolumn{1}{c}{(6)}\\
			&\multicolumn{1}{c}{OLS}&\multicolumn{1}{c}{OLS}&\multicolumn{1}{c}{OLS}&\multicolumn{1}{c}{OLS}&\multicolumn{1}{c}{OLS}&\multicolumn{1}{c}{OLS}\\
			&\multicolumn{1}{c}{Grade}&\multicolumn{1}{c}{Grade}&\multicolumn{1}{c}{Grade}&\multicolumn{1}{c}{Grade}&\multicolumn{1}{c}{Grade}&\multicolumn{1}{c}{Grade}\\
			\hline
			
			Order           &  -0.0048*  &  -0.0001   &   0.0010   &            &            &            \\
			& (0.0025)   & (0.0018)   & (0.0014)   &            &            &            \\
			Program=Bachelor $2^{nd}$ year        &            &   0.6093***&            &            &   0.6136***&            \\
			&            & (0.1775)   &            &            & (0.1738)   &            \\
			Program=Bachelor $3^{rd}$ year        &            &   0.7111***&            &            &   0.7136***&            \\
			&            & (0.1550)   &            &            & (0.1532)   &            \\
			Program=Master       &            &   0.9875***&            &            &   0.9921***&            \\
			&            & (0.1559)   &            &            & (0.1517)   &            \\
			9-10am     &            &            &            &   0.0246   &  -0.0084   &   0.0173   \\
			&            &            &            & (0.0278)   & (0.0265)   & (0.0230)   \\
			11am-12pm   &            &            &            &   0.0975** &   0.0764** &   0.0609** \\
			&            &            &            & (0.0403)   & (0.0350)   & (0.0261)   \\
			1-2pm     &            &            &            &   0.0183   &  -0.0076   &  -0.0262   \\
			&            &            &            & (0.0490)   & (0.0442)   & (0.0329)   \\
			3-4pm     &            &            &            &   0.0494   &   0.0512   &   0.0204   \\
			&            &            &            & (0.0422)   & (0.0366)   & (0.0198)   \\
			5-6pm     &            &            &            &   0.0218   &   0.0668*  &   0.0138   \\
			&            &            &            & (0.0529)   & (0.0366)   & (0.0428)   \\
			\hline
			Fixed Effects & No & No & Yes & No & No & Yes \\
			Observations    &    14658   &    14658   &    13776   &    14658   &    14658   &    13776   \\
			\(R^{2}\)       &    0.001   &    0.113   &    0.557   &    0.001   &    0.115   &    0.557   \\
			\hline\hline
			\multicolumn{7}{c}{\footnotesize Note: Standard errors in parentheses. * p<0.1, ** p<0.05, *** p<0.01.}\\
			\multicolumn{7}{c}{\footnotesize Program: Bachelor 1st year is the reference category. }\\
			\multicolumn{7}{c}{\footnotesize Fixed effects estimations: dropped 882 singleton observations.} \\ \hline
		\end{tabular}
	}
\end{table}
%\vspace{-0.4 cm}
%\begin{minipage}{0.46\textwidth}
%\tiny Note: Standard errors in parentheses. Program: Bachelor 1st year is the reference category. * p<0.1, ** p<0.05, *** p<0.01. \\

%\end{minipage}

	\begin{comment}
		\input{old_reg_tables/regorder}
	\end{comment}
	
	\section*{Results}
	
	We first examine whether our data provides evidence for the presence of mental depletion in our context of interest. If professors get increasingly exhausted when a series of students pass the same exam on a given day, one would expect that the grades systematically change with the running order. Descriptive Figure \ref{fig:figure1}, Panel A, represents the average grade by order of scheduling. For each position in the running order with at least 100 observations, we take the average grade.  The average grade per position in the running order varies from 4.5 to nearly 4.9, with a downward trend, as the order increases, more particularly starting from the $22^{nd}$ passage onwards. The statistical significance of this negative trend is confirmed by the regression analysis reported in Column (1) of Table \ref{tab:T1}. This first estimation does not include any control variables or fixed effects and therefore simply established the raw association between the grade given by the professor and the student's running order. According to this regression, a one-unit increase in the running order decreases the grade by $0.0048$ ($p=0.058$).
	
	%Whereas the effect size of this decline may seem rather small when considering the full range of grades from 1 to 6, it is important to take into account that the vast majority of grades (86\%) falls into the range from 4 (passing grade) to 6 (highest grade). Within this most relevant range of the grade scale the decline is substantial. 
	If our study were to suffer from similar data limitations as past work, one might therefore be tempted to draw the (erroneous) conclusion that a relevant degree of decision fatigue is present in our sample as well. However, our data allows us to dig deeper. In the next step we split the sample by different programs in which the exams took place (first-year, second-year and third year Bachelor as well as Master). Figure \ref{fig:figure1}, Panel B, displays the association between grades and running order for each program separately. This way of presenting the data reveals two interesting insights. First, grades in the Bachelor years tend to be lower than those in the Master program (most pronouncedly so for the first-year Bachelor exams). Second, there is no systematically negative association between grades and the running order at the program level. These observations suggest that it might be important to control for program effects. And indeed, Column (2) of Table \ref{tab:T1} confirms that the negative association between grades and the running order disappears once we we add program fixed effects to our regression. This null result remains robust if we further exploit the richness of our data. Column (3) reports the results of specifications that include professor, exam, student and date fixed effects. Columns (4)-(6) present the results of analogous specifications as in Columns (1)-(3), but focusing instead of the passing order variable on a set of dummy variables for specific time pariods across the day. Overall, we do not see a systematic association between grades and the running order or specific times in this Table.
	
	So, where does the downward trend observed in Figure \ref{fig:figure1}, Panel A, come from? A simple analysis of the distribution of observations across the running order reveals that our data set contains a substantially higher proportion of grades form Bachelor programs at higher positions in the running order (see Figure \ref{fig:figure1}, Panel C). Accordingly, the seemingly striking pattern of Figure \ref{fig:figure1}, Panel A, is simply a spurious artifact of composition bias. %In fact, we can even push this argument one level further. The only case in which grades seem to be lower for higher positions in the running order is the first-year Bachelor (see Figure \ref{fig:figure1}, Panel B). If we analyze the distribution of observations at the exam level, we find again evidence for a composition effect, because the decrease for high positions in the running order is driven by a set of particularly large courses that tend to have lower grades in general.
	
	%\begin{figure}[h!]
	%\centering
	%\includegraphics[width=1\linewidth]{new_figures/Fig2A.pdf}
	%\caption{Mean grades before and after the break}
	%\label{fig:figure2a}
	%\end{figure}
	
\begin{table}[H]
	\centering
	\caption{Effects of running order before and after breaks\label{tab:breaks}}
	\adjustbox{max width=0.5\textwidth}{
		\begin{tabular}{lccc}
			\hline\hline
			%&\multicolumn{1}{c}{(1)}&\multicolumn{1}{c}{(2)}&\multicolumn{1}{c}{(3)}\\
			%&\multicolumn{1}{c}{Grade}&\multicolumn{1}{c}{Grade}&\multicolumn{1}{c}{Grade}\\
			&(1)&(2)&(3)\\
			&OLS&OLS&FE\\
			&Grade&Grade&Grade\\
			\hline
			
			Order           &   0.0084** &   0.0140***&   0.0089***\\
			& (0.0036)   & (0.0029)   & (0.0023)   \\
			After Break  &   0.2206*  &   0.2283***&   0.0634   \\
			& (0.1132)   & (0.0791)   & (0.0758)   \\
			After Break $\times$ Order&  -0.0177***&  -0.0216***&  -0.0103** \\
			& (0.0055)   & (0.0043)   & (0.0044)   \\
			No Lunch Break  &   0.1600***&   0.0857** &   0.0032   \\
			& (0.0556)   & (0.0378)   & (0.0206)   \\
			Program=Bachelor $2^{nd}$ year        &            &   0.6052***&            \\
			&            & (0.1664)   &            \\
			Program=Bachelor $3^{rd}$ year     &            &   0.7292***&            \\
			&            & (0.1457)   &            \\
			Program=Master     &            &   0.9813***&            \\
			&            & (0.1473)   &            \\
			\hline
			Fixed Effects & No & No & Yes\\
			Observations    &    14658   &    14658   &    13776   \\
			\(R^{2}\)       &    0.009   &    0.118   &    0.558   \\
			\hline\hline
			\multicolumn{4}{c}{\footnotesize Note: Standard errors in parentheses. * p<0.1, ** p<0.05, *** p<0.01.} \\
			\multicolumn{4}{c}{\footnotesize Program: Bachelor 1st year is the reference category. } \\
			\multicolumn{4}{c}{\footnotesize Fixed effects estimations: dropped of singleton observations.} \\ \hline
		\end{tabular}
	}
\end{table}
%\vspace{-0.4 cm}
%\begin{minipage}{0.56\textwidth}\tiny Note: Standard errors in parentheses. Program: Bachelor 1st year is the reference category. * p<0.1, ** p<0.05, *** p<0.01. \\
%Fixed effects specification: dropped singleton observations.
%\end{minipage}

	The above evidence indicates that, overall, mental depletion does not seem to be a major concern in our context. To provide further support for this interpretation, we also examine the effects of breaks, which play a central role for the interpretation of mechanisms in the previous literature 
	\cite{danziger2011extraneous}. Figure \ref{fig:figure1}, Panel D displays graphically the average grade by order of scheduling before and after the break. The corresponding Table \ref{tab:breaks} presents a regression table investigating the analogous question. In strong contrast to the findings for the judges in Israel \cite{danziger2011extraneous}, we do not observe a decreasing trend in the running order that gets reset after the break. In fact, grades tend to increase in the running order before the break and then tend to decrease after the break. This pattern is clearly incompatible with the hunger explanation.
	
	%XXX We need to explain how this graph was constructed. I am not sure which observations are included and how the distance to the break is determined. Maybe we should only include observations with a sufficient number of obs before and after the break (otherwise the trends might be heavily influenced by short exams with a break, not sure how many of those we have but to be checked). XXX 
	
	%However, once we control professor fixed effects in column 2, the coefficient change sign. Indeed, we see a positive effect of an increase in the position in the running order on the grade. The coefficient becomes even more statistically and economically significant once we add the exam fixed effects (column 3); an increase of one in the running order increases the grade by $0.0039$  holding everything else constant. However, once we take into account the student fixed effects, the relation become insignificant (column 4). Adding semester-year or date fixed effects don't change the null effect found in column 4 (columns 5 and 6).
	
	Another finding identified in the previous literature \cite{chen2016decision} is the presence of \textit{negative} auto-correlation. This effect is typically interpreted as a form of ``gambler fallacy''. We also investigate the extent to which such an effect can be found in our data set. To do so, we determine the impact of the grades in the three most recent exams on the grade in the current exam. We find that--if anything--our data is characterized by \textit{positive} auto-correlation. %In Table \ref{tab:regorder_lag}, we find a similar pattern for the \textit{Order} coefficient for the linear regression (column 1), but as soon as we add professors fixed, column 2, the \textit{Order} coefficient become not significant, converging to the idea that the running order does not impact significantly the grade. 
	The auto-correlation variable in Table \ref{tab:T2}, \textit{Average Last 3 Lag Grades}, is statistically significant at the 5 percent level and positive from column 1 to 4; suggesting higher previous grades are associated with higher current grades. However, we note that the auto-correlation coefficient is becoming less and less quantitatively significant as we add fixed effects and the estimated coefficients are getting smaller.

\begin{table}[H]
	\centering
	\caption{Main results on auto-correlation of grades\label{tab:T2}}
	\adjustbox{max width=0.5\textwidth}{
		\begin{tabular}{l*{5}{c}}
			\hline\hline
			
			&\multicolumn{1}{c}{(1)}&\multicolumn{1}{c}{(2)}&\multicolumn{1}{c}{(3)}&\multicolumn{1}{c}{(4)}&\multicolumn{1}{c}{(5)}\\
			&\multicolumn{1}{c}{OLS}&\multicolumn{1}{c}{OLS}&\multicolumn{1}{c}{FE}&\multicolumn{1}{c}{OLS}&\multicolumn{1}{c}{OLS}\\
			&\multicolumn{1}{c}{Grade}&\multicolumn{1}{c}{Grade}&\multicolumn{1}{c}{Grade}&\multicolumn{1}{c}{Grade}&\multicolumn{1}{c}{Grade}\\
			\hline
			Average Last 3 Lag Grades&   0.4644***&   0.2949***&   0.0515*  &            &            \\
			& (0.0371)   & (0.0325)   & (0.0286)   &            &            \\
			Order           &  -0.0051***&  -0.0032** &  -0.0014   &  -0.0000   &  -0.0039   \\
			& (0.0014)   & (0.0014)   & (0.0018)   & (0.0018)   & (0.0026)   \\
			Program=Bachelor $2^{nd}$ year       &            &   0.4458***&            &            &            \\
			&            & (0.1298)   &            &            &            \\
			Program=Bachelor $3^{rd}$ year       &            &   0.5129***&            &            &            \\
			&            & (0.1118)   &            &            &            \\
			Program=Master       &            &   0.7218***&            &            &            \\
			&            & (0.1118)   &            &            &            \\
			Lag Grade       &            &            &            &   0.0507***&            \\
			&            &            &            & (0.0125)   &            \\
			Lag 5 Previous Grades    &            &            &            &            &   0.0675   \\
			&            &            &            &            & (0.0428)   \\
			\hline
			Fixed Effects & No & No & Yes & No & No\\
			Observations    &    11224   &    11224   &    10348   &    12459   &     8407   \\
			\(R^{2}\)       &    0.110   &    0.157   &    0.569   &    0.563   &    0.568   \\
			\hline\hline
			\multicolumn{6}{c}{\footnotesize Note: Standard errors in parentheses. * p<0.1, ** p<0.05, *** p<0.01.} \\
			\multicolumn{6}{c}{\footnotesize Program: Bachelor 1st year is the reference category. } \\
			\multicolumn{6}{c}{\footnotesize Fixed effects estimations: dropped of singleton observations.} \\ 
			\multicolumn{6}{c}{\footnotesize Inclusion of Lag Variables: dropped of some observations.} \\ \hline
		\end{tabular}
		
	}
\end{table}
%\vspace{-0.4 cm}
%\begin{minipage}{0.46\textwidth}\tiny Note: Standard errors in parentheses. Program: Bachelor 1st year is the reference category. * p<0.1, ** p<0.05, *** p<0.01. \\
%Fixed effects specification: dropped singleton observations.\end{minipage}
	
	In the appendix we include also additional regression results, where we re-run the main (most demanding) baseline regression specifications to which we add further control variables, such as whether the professor-student combination featured two females or two males (with all other combinations being the reference category), whether the exam took place on zoom, whether it was the first or last exam of the day, and whether there was no lunch break. As displayed in Table \ref{tab:AT2}, the conclusions of our analysis are unchanged when including these additional variables. 
	
	\section*{Discussion}
	
	There is ample evidence that spectacular and surprising results have higher chances of being published in high-impact journals than null-results. This is very problematic, because such a publication bias can seriously flaw policy advice. Regarding the impact of decision scheduling on choice sequences, several well-published and influential articles find evidence for decision fatigue and biased decisions. Researchers have argued that their findings may only represent the ``tip of the iceberg'', because the underlying mechanisms at work (chemical processes (low glucose levels) or psychological channels) appear to be quite general.
	
	To make sure that policy advice is not based on overgeneralized or misleading interpretations, it is  extremely important to study the impact of decision scheduling across a wide range of settings, especially with data that allow researchers to circumvent the limitations that potentially threaten the clean identification of causal effects. In our study, the exogeneity of the scheduling and the large sample size allows us to control for student and professor characteristics and filter out the key potential confounding factors. In this very demanding statistical setting, we find the null result that the scheduling order does not affect the grading decision. This finding is in line with the (reassuring) policy conclusion that the previous results on decision fatigue may not extend to different contexts.

	%\showmatmethods{} % Display the Materials and Methods section
	
	%\acknow{We are very grateful to Daisy Riccio Buqi for the outstanding help with the data preparation. We also thank the Ethics Committee of the Faculty of Business and Economics (HEC) of the University of Lausanne for their helpful comments.}
	
	%\showacknow{} % Display the acknowledgments section
	
	% Bibliography
	\bibliographystyle{abbrv}
	\bibliography{pnas-sample.bbl}
	%\clearpage
	\newpage
	
	\setcounter{table}{0}
	\renewcommand{\thetable}{A\arabic{table}}
	
	\section*{Appendix}
	
	Below are included two Appendix Tables referred to in the main text. In particular, Table \ref{tab:AT1} presents the descriptive summary statistics, while Table \ref{tab:AT2} reproduces the most demanding specifications of the baseline tables, while adding further control variables.

	\begin{table}[htbp]\centering
		\caption{\label{tab:AT1} 
			\textbf{} }\adjustbox{max width=0.5\textwidth}{\begin{tabular} {@{} l r r r r r @{}} \\ \hline \hline
				\textbf{Variable } & \textbf{         N} & \textbf{      Mean} & \textbf{        SD} & \textbf{       
					Min} & \textbf{       Max} \\
				\hline
				Grade  &      14658 &        4.8 &        .93 &          1 &          6 \\
				Order  &      14658 &        9.1 &        6.6 &          1 &         32 \\
				Program=Bachelor $1^{st}$ year         &      14658 &        .11 &        .32 &          0 &          1 \\
				Program=Bachelor $2^{nd}$ year          &      14658 &        .18 &        .38 &          0 &          1 \\
				Program=Bachelor $3^{rd}$ year          &      14658 &        .15 &        .36 &          0 &          1 \\
				Program=Master        &      14658 &        .56 &         .5 &          0 &          1 \\
				9-10am  &      14658 &        .31 &        .46 &          0 &          1 \\
				11am-12pm  &      14658 &        .13 &        .33 &          0 &          1 \\
				1-2pm  &      14658 &        .16 &        .37 &          0 &          1 \\
				3-4pm &      14658 &        .24 &        .43 &          0 &          1 \\
				5-6pm &      14658 &       .055 &        .23 &          0 &          1 \\
				First exam  &      14658 &       .082 &        .27 &          0 &          1 \\
				Last exam  &      14658 &       .081 &        .27 &          0 &          1 \\
				No break  &      14658 &         .5 &         .5 &          0 &          1 \\
				On zoom  &      14658 &        .25 &        .44 &          0 &          1 \\
				Both females  &      14658 &        .16 &        .36 &          0 &          1 \\
				Both males &      14658 &        .28 &        .45 &          0 &          1 \\
				\hline \hline 
				%\multicolumn{6}{@{}l}{\footnotesize{\emph{Source:} 2018_2021_v5.dta}}
			\end{tabular}
		}
	\end{table}

	\begin{table}[H]
		\centering
		\caption{Further results on passing order and times\label{tab:AT2}}
		\adjustbox{max width=0.5\textwidth}{
			\begin{tabular}{l*{4}{c}}
				\hline\hline
				&\multicolumn{1}{c}{(1)}&\multicolumn{1}{c}{(2)}&\multicolumn{1}{c}{(3)}&\multicolumn{1}{c}{(4)}\\
				&\multicolumn{1}{c}{FE}&\multicolumn{1}{c}{FE}&\multicolumn{1}{c}{FE}&\multicolumn{1}{c}{FE}\\
				&\multicolumn{1}{c}{Grade}&\multicolumn{1}{c}{Grade}&\multicolumn{1}{c}{Grade}&\multicolumn{1}{c}{Grade}\\
				\hline
				Order           &  -0.0002   &            &  -0.0007   &            \\
				& (0.0018)   &            & (0.0020)   &            \\
				First exam of the day&  -0.0697** &  -0.0610** &     &      \\
				& (0.0339)   & (0.0301)   &       &        \\
				Last exam of the day&   0.0508   &   0.0460   &   0.0404   &   0.0322   \\
				& (0.0350)   & (0.0349)   & (0.0387)   & (0.0395)   \\
				No lunch break  &   0.0143   &   0.0126   &   0.0403   &   0.0371   \\
				& (0.0214)   & (0.0207)   & (0.0296)   & (0.0262)   \\
				On zoom            &  -0.0345   &  -0.0343   &  -0.0197   &  -0.0184   \\
				& (0.0698)   & (0.0721)   & (0.0659)   & (0.0672)   \\
				Both female     &  -0.3277*  &  -0.3353*  &  -0.0844   &  -0.0980   \\
				& (0.1910)   & (0.1888)   & (0.2121)   & (0.2124)   \\
				Both male      &   0.3085   &   0.3155   &   0.0778   &   0.0897   \\
				& (0.1940)   & (0.1925)   & (0.2109)   & (0.2116)   \\
				9-10am     &            &   0.0025   &            &   0.0359   \\
				&            & (0.0257)   &            & (0.0422)   \\
				11am-12pm     &            &   0.0373   &            &   0.0831*  \\
				&            & (0.0304)   &            & (0.0481)   \\
				1-2pm    &            &  -0.0371   &            &   0.0008   \\
				&            & (0.0349)   &            & (0.0506)   \\
				3-4pm     &            &   0.0004   &            &   0.0403   \\
				&            & (0.0237)   &            & (0.0446)   \\
				5-6pm     &            &  -0.0213   &            &   0.0206   \\
				&            & (0.0488)   &            & (0.0537)   \\
				Average Last 3 Lag Grades&            &            &   0.0517*  &   0.0509*  \\
				&            &            & (0.0286)   & (0.0280)   \\
				\hline
				Fixed Effects & Yes & Yes & Yes & Yes\\
				Observations    &    13776   &    13776   &    10348   &    10348   \\
				\(R^{2}\)       &    0.557   &    0.558   &    0.569   &    0.569   \\
				\hline\hline
				\multicolumn{5}{c}{\footnotesize Note: Standard errors in parentheses. * p<0.1, ** p<0.05, *** p<0.01.}\\
				\multicolumn{5}{c}{\footnotesize Program: Bachelor 1st year is the reference category. }\\ 
				\multicolumn{5}{c}{\footnotesize Inclusion of Lag Variables: dropped of some observations.} \\ \hline
			\end{tabular}
			
		}
	\end{table}
\end{document}